\documentclass[aps, prb, preprint, showkeys, showpacs]{revtex4}
\usepackage{amsmath}
\usepackage{graphicx}
\usepackage{amsfonts}

\renewcommand{\L}{\mathcal{L}}
\newcommand{\vf}{v_\text{F}}
\newcommand{\D}{\mathcal{D}}
\newcommand{\Z}{\mathcal{Z}}
\newcommand{\N}{\mathcal{N}}
\renewcommand{\H}{\mathcal{H}}

\newcommand{\unmedio}{{\scriptstyle\frac{1}{2}}}
\newcommand{\eff}{_{\text{eff}}}

\newcommand{\tr}{\operatorname{tr}}
\newcommand{\psiB}{\bar{\psi}}

\newcommand{\chiB}{\bar{\chi}}

\begin{document}

\title{A non covariant fermionic determinant and its connection to Luttinger systems}
\author{An\'{\i}bal Iucci and Carlos Na\'on}
\email{iucci@fisica.unlp.edu.ar; naon@fisica.unlp.edu.ar}

\affiliation{Instituto de F\'{\i}sica La Plata. Departamento de
F\'{\i}sica, Facultad de Ciencias Exactas, Universidad Nacional de
La Plata. CC 67, 1900 La Plata, Argentina.\\ Consejo Nacional de
Investigaciones Cient\'{\i}ficas y T\'ecnicas, Argentina.}

\begin{abstract}
We consider a fermionic determinant associated to a non covariant
Quantum Field Theory used to describe a non relativistic system in
$(1+1)$ dimensions. By exploiting the freedom that arises when
Lorentz invariance is not mandatory, we determine the heat-kernel
regulating operator so as to reproduce the correct dispersion
relations of the bosonic excitations. We also derive the
Hamiltonian of the functionally  bosonized model and the
corresponding currents. In this way we were able to establish the
precise heat-kernel regularization that yields complete agreement
between the path-integral and operational approaches to the
bosonization of the Tomonaga-Luttinger model.
\end{abstract}



\maketitle
\section{Introduction}
 Fermionic determinants play a central role in modern
formulations of Quantum Field Theories (QFT's). As it is
well-known, they naturally arise when considering generating
functionals associated to fermion fields in the path-integral
approach \cite{Ramond}. In the last twenty years it has been
specially fruitful the study of fermionic determinants in (1+1)
dimensions. Fujikawa's observation concerning the non triviality
of the Jacobian associated to chiral changes in the fermionic
functional integration measure \cite{Fujikawa}, when specialized
to the (1+1)-dimensional case, led to significant advances in our
understanding of paradigmatic ``toy-models" such as
two-dimensional quantum electrodynamics ($QED_{2}$), the Thirring
model, and their non-Abelian versions \cite{toy}. In fact, a
functional bosonization technique was developed based on an
adequate treatment of the fermionic determinant \cite{Stone}. The
crucial point is that the above mentioned Jacobian needs a
regularization. For gauge theories with Dirac fermions one is
naturally led to consider a regularization scheme that preserves
gauge invariance. On the other hand, when the vector fields which
are present in the theory are just auxiliary fields (usually
introduced through a Hubbard-Stratonovich transformation), one can
choose a more general regulator \cite{Rubin}$\,$\cite{CS}. The
Thirring model \cite{Thirring} and the chiral Schwinger model
\cite{CSM} are examples in which regularization ambiguities take
place.

The issue of the regularization of the Fujikawa Jacobian, its
relation to local counterterms, and its role in the analysis of
quantum anomalies has been extensively examined in the literature
\cite{Fujikawa2}. In all cases the models under study are
relativistic QFT's, i.e. Lorentz covariant theories. However, in
certain relevant situations one is interested in non covariant
field theories. This is the case in the analysis of one
dimensional (1D) electronic systems which can be studied through
the g-ology model \cite{Solyom}, a theory with four coupling
constants $g_1$,...$g_4$ associated to different scattering
processes (for $g_1=g_3=0$ it reduces to a low-energy model for
electrons known as the Tomonaga-Luttinger model \cite{reviews}).
In this context, the existence of a functional bosonization,
alternative to the usual operator approach, was first suggested by
Fogebdy \cite{Fogebdy} and further elaborated by Lee and Chen
\cite{LC}. The explicit connection between the functional
bosonization leading to an effective action describing the
dynamics of bosonic collective excitations and the Fujikawa
Jacobian was first established in ref. \onlinecite{Naon}. But even
in this case a covariant regularization, borrowed from
relativistic field theory, was employed. As a result, the general
expressions for the dispersion relations of the bosonic modes, in
terms of the couplings $g_{2}$ and $g_{4}$ of the
Tomonaga-Luttinger model, did not agree with the ones obtained
through conventional, operational bosonization.

In this work we show that the origin of this disagreement is in
the type of regularization chosen to compute the Fujikawa
Jacobian. In Section II we present the model and express its
generating functional in terms of a fermionic determinant. In
Section III, in order to clarify the discussion, we start by
sketching the main steps of the decoupling approach to
bosonization and the results obtained using a standard Lorentz
invariant regularization. We then include two subsections A and B
where we present two different non-covariant regularizations, the
point-splitting and the heat-kernel methods, respectively. In this
last case we determine the precise form of the regulating
heat-kernel operator needed to obtain the right answer for the
dispersion relations. In Section IV we show how to derive, in our
functional bosonization framework, the bosonic Hamiltonian and the
corresponding bosonized currents. Finally we briefly discussed the
issue of current conservation. In Section V we gather our results
and conclusions.

\section{The model and the Fermionic Determinant}

We will consider a non covariant version of the Thirring model
defined by the following Euclidean lagrangian

\begin{equation}
\L=\psiB i \partial\!\!\!\slash \psi - \frac{g^2}{2}
V_{(\mu)}j_\mu j_\mu,
\end{equation}
where $V_0$ and $V_1$ are the coupling constants and the
derivatives are redefined in order to include the Fermi velocity,

\begin{align}
\partial_0=&\frac{\partial}{\partial x_0}\\
\partial_1=&\vf\frac{\partial}{\partial x_1}.
\end{align}
Note that $\vf$ plays the role of the light velocity in QFT, which
is usually taken as unit. For $\vf=1$ and $V_0=V_1=1$ one has the
usual Thirring model (the constant $g^2$ is included to facilitate
comparison with the Lorentz invariant results). The fermionic
current is defined as

\begin{equation}
j_\mu=\psiB\gamma_\mu\psi,
\end{equation}
which satisfies the classical conservation law

\begin{equation}
\partial_\mu j_\mu=0.
\end{equation}

The generating functional is

\begin{equation}
\Z[S]=\int\D\psiB\D\psi\,\exp\left[-\int d^2x(\L+j_\mu
S_\mu)\right].
\end{equation}

By means of a Hubbard-Stratonovich transformation, it can be put
in the form

\begin{equation}\label{generatingFunctional}
\Z[S]=\N\int\D A_\mu \det D\!\!\!\!\slash\;[A]
\,\exp\left[-\frac{1}{2g^2}\int
d^2x\,d^2y\,V^{-1}_{(\mu)}(x-y)(gA_\mu
-S_\mu)(x)(gA_\mu-S_\mu)(y)\right],
\end{equation}
where

\begin{equation}
 D\!\!\!\!\slash\;[A]=i \partial\!\!\!\slash + g A\!\!\!\slash ,
\end{equation}
and

\begin{equation}
\det  D\!\!\!\!\slash\;[A]=\int\D\psiB\D\psi\,\exp\left[-\int
d^2x\,\psiB  D\!\!\!\!\slash\;[A]\psi\right].
\end{equation}

\section{Decoupling approach to bosonization}

Having expressed the generating functional in terms of a fermionic
determinant we shall now sketch the decoupling method which is at
the root of the functional approach to bosonization pioneered in
ref. \onlinecite{toy} (see also ref. \onlinecite{Stone}). In
$(1+1)$ space-time the vector field $A_\mu$ can be decomposed in
transverse and longitudinal pieces:

\begin{equation}\label{AphietaRelation}
A_\mu = -(1/g) (\epsilon_{\mu\nu}
\partial_\nu\phi-\partial_\mu\eta),
\end{equation}
where $\phi$ and $\eta$ are scalar fields. Let us note that if we
perform the following transformation in the fermionic fields

\begin{align}
\psi=&e^{t[\gamma_5\phi+i\eta]}\chi\\
\psiB=&e^{t[\gamma_5\phi-i\eta]}\chiB,
\end{align}
then the fermionic lagrangian density changes as

\begin{equation}
\psiB D\!\!\!\!\slash\;[A]\psi=\chiB D\!\!\!\!\slash_t[A]\chi
\end{equation}
where

\begin{equation}
 D\!\!\!\!\slash_t[A]= D\!\!\!\!\slash\;[(1-t)A].
\end{equation}

As first observed by Fujikawa \cite{Fujikawa}, the Jacobian
associated to the above change in the path integration measure is
not trivial, it depends on the fields $\phi$ and $\eta$:

\begin{equation}
\det(i
\partial\!\!\!\slash +gA\!\!\!\slash )=J[\phi,\eta;t]\det(i \partial\!\!\!\slash +g(1-t)A\!\!\!\slash ).
\end{equation}
Note that for $t=1$ the fermionic and bosonic degrees of freedom become completely
decoupled. It can be shown that

\begin{equation}\label{jacobian}
J[\phi,\eta;1]\equiv J =\exp\left[-\int_0^1\omega(t)\right]
\end{equation}
with

\begin{equation}\label{omega}
\omega(t)=-\tr D\!\!\!\!\slash_t[A]^{-1}\,g\,A\!\!\!\slash
=-\lim_{y\rightarrow x}\tr^D\int d^2x\,
D\!\!\!\!\slash_t[A]^{-1}(x,y)\, g\,A\!\!\!\slash (x),
\end{equation}
where $\tr^D$ means the trace in the Dirac space. All these
formulae close a deep analogy with the ones corresponding to a
covariant QFT. Actually the only difference between them is the
presence of $\vf$ instead of the velocity of light, but this has
non trivial implications. This last equation has to be
regularized, otherwise divergences appear, as it is obvious by
taking the $y\rightarrow x$ limit. In QFT any acceptable
regularizing method has to be Lorentz invariant. In the present
case we do not have that limitation for two reasons: i) Lorentz
invariance is broken from the beginning since we are in a
non-relativistic theory; ii) There is a remaining covariance, i.
e. the theory without interactions is invariant with respect to
the Lorentz group where the velocity of light has been replaced by
$\vf$, but this is an artificial symmetry and there is no reason
to respect it. Moreover, by taking $V_0\neq V_1$ ($g_2\neq g_4$)
not even this symmetry is present. Before taking advantage of the
freedom arising from the absence of covariance, it could be
instructive to recall the results previously obtained by choosing
a Lorentz invariant regularization \cite{Naon}. Using a regulator
of the form

\begin{equation}
\left( D\!\!\!\!\slash_t[A]\, D\!\!\!\!\slash_t[A]^{\dagger} +
 D\!\!\!\!\slash_t[A]^{\dagger}\, D\!\!\!\!\slash_t[A]\right)/2,
\end{equation}
which was first proposed by Fujikawa in his analysis of covariant
and consistent anomalies \cite{Fujikawa3}, one gets

\begin{equation}
J_\text{cov}=\exp\left\{-\frac{a}{2\pi\vf}\int
d^2x\,\left[(\partial_1\phi)^2+(\partial_0\phi)^2\right]\right\},
\end{equation}
where $a$ is a parameter related to possible regularization
ambiguities. For $a=1$ one has a gauge invariant regularization.
Although the Thirring model does not possess local gauge
invariance, in the present context we are mainly interested in
Lorentz invariance and we can then set $a=1$ without loss of
generality. Inserting the above Jacobian in the generating
functional, absorbing the free fermion determinant, which results
from the decoupling procedure, in a normalization factor and
expressing $A_\mu$ in terms of $\phi$ and $\eta$ according to eq.
(\ref{AphietaRelation}), one obtains a bosonized action. In the
condensed-matter context these bosonic degrees of freedom are
interpreted as fields associated to charge-density oscillations.
From this bosonic action derived through a covariance-preserving
regularization one can easily compute the corresponding dispersion
relation:

\begin{equation}
p_0^2 + v_\text{cov}^2\,p_1^2=0
\end{equation}
where

\begin{equation}
v_\text{cov}^2=\vf^2\frac{\left(\vf+\frac{g^2 V_0}{\pi}\right)}
{\left(\vf+\frac{g^2 V_1}{\pi}\right)}.
\end{equation}
Let us stress that only for $V_1=0$ this velocity agrees with the
value obtained by using operational bosonization, which reads:

\begin{equation}\label{velocity}
v^2=\left(\vf-\frac{V_1 g^2}{\pi}\right)\left(\vf+\frac{V_0
g^2}{\pi}\right).
\end{equation}

In the following we will describe two different methods to
regularize the Jacobian that do not preserve Lorentz invariance
and allow to obtain an effective bosonic action which leads to the
right answer for the dispersion relation.

\subsection{Point-splitting Method}

As it is well-known, the point-splitting regularization method
breaks Lorentz invariance explicitly. It consists in a
prescription for taking the $y\rightarrow x$ limit mentioned
before by defining

\begin{equation}\label{defPointSplitting}
\lim_{y\rightarrow x}
D\!\!\!\!\slash_t[A]^{-1}(x,y)=\unmedio(\lim_{\epsilon\rightarrow
0^+}+\lim_{\epsilon\rightarrow 0^-})
D\!\!\!\!\slash_t[A]^{-1}(x_0,x_1;x_0,x_1+\epsilon),
\end{equation}
i.e. by taking a symmetric limit in the space variable. We need
then the Green's function of the Dirac operator, which satisfies

\begin{equation}
 D\!\!\!\!\slash_t[A]_x D\!\!\!\!\slash_t[A]^{-1}(x,y)=\delta^2(x-y)
\end{equation}

As usual, we propose the ansatz

\begin{equation}
 D\!\!\!\!\slash
_t[A]^{-1}(x,y)=e^{(1-t)\left[\gamma_5\phi(x)+i\eta(x)\right]}\,
G_0(x,y)\, e^{(1-t)\left[\gamma_5\phi(y)-i\eta(y)\right]},
\end{equation}
where $G_0$ is the Green´s function of the free Dirac operator:

\begin{equation}
i \partial\!\!\!\slash _x G_0(x,y)=\delta^2(x-y).
\end{equation}

With this recipe, we find for equation (\ref{defPointSplitting})
the following result

\begin{equation}
\lim_{y\rightarrow x} D\!\!\!\!\slash
_t[A]^{-1}(x,y)=-\frac{i}{2\pi\vf}(1-t)\,\gamma_1
\partial_1 \left[\gamma_5\phi(x)-i\eta(x)\right]
\end{equation}
and then, the Jacobian (eqs. (\ref{jacobian}) and (\ref{omega}))
is given by

\begin{equation}\label{resultJacobian}
J=\exp\left\{-\frac{1}{2\pi\vf}\int
d^2x\,\left[(\partial_1\phi)^2-(\partial_1\eta)^2-2\partial_1\phi\partial_0\eta\right]\right\}.
\end{equation}
The vacuum functional can then be written as

\begin{equation}
\Z[S=0]= \N \int\D\phi\D\eta\,e^{-S\eff}
\end{equation}
where $\N$ is a normalization factor that includes the free
fermion (interaction independent) determinant. We have also
defined $S\eff$, which in momentum space takes the form

\begin{equation}
S\eff=\int
\frac{d^2p}{(2\pi)^2}\,\left[\phi(p)A\phi(-p)+\eta(p)B\eta(-p)+2\phi(p)C\eta(-p)\right].
\end{equation}
with

\begin{align}
A=&\vf^2p_1^2\left(\frac{1}{2g^2V_0}+\frac{1}{2\pi\vf}\right)+\frac{p_0^2}{2g^2V_1}\\
B=&\vf^2p_1^2\left(\frac{1}{2g^2V_1}-\frac{1}{2\pi\vf}\right)+\frac{p_0^2}{2g^2V_0}\\
C=&p_1p_0\vf\left(\frac{1}{2g^2V_1}-\frac{1}{2g^2V_0}-\frac{1}{2\pi\vf}\right)
\end{align}

The physical content of the model can be extracted from $S\eff$
which describes the dynamics of the collective modes of the
system. When the original fermionic model is related to the
Tomonaga-Luttinger model used to study 1D electronic systems
\cite{reviews}, these collective excitations correspond to
charge-density oscillations (plasmons). Their dispersion relation
can be obtained as the zeros of the determinant of the matrix
\begin{equation}\label{Mmatrix}
\begin{pmatrix}
      A & C \\
      C & B \\
\end{pmatrix}.
\end{equation}
The result is
\begin{equation}
p_0^2 + v^2\,p_1^2=0
\end{equation}
where $v$ is the renormalized velocity of the charge-density modes
given by eq. (\ref{velocity}).

\subsection{Heat-kernel Method}

Another popular way of dealing with the regularization of
fermionic determinants is the heat-kernel method
\cite{Fujikawa}$\,$\cite{heat}. In this scheme $J$ is regulated by
inserting an operator of the form $e^{-R/M^2}$, $R$ is a positive
definite operator and $M$ is a mass-like parameter which is kept
fixed in the intermediate computations. The limit
$M^2\rightarrow\infty$ is taken at the end. Again, let us
emphasize that in standard QFT contexts the regulating operator
$R$ can be chosen among all operators compatible with Lorentz
invariance (let aside, for the moment, any other possible
symmetries), for instance $R= D\!\!\!\!\slash _t[A]^2$. Here we do
not have that limitation and our purpose is to find the precise
form of $R$ that leads to an effective action containing the
desired dispersion relation.

We start by rewriting equation (\ref{omega}) as

\begin{equation}\label{omega2}
\omega(t)=\tr\left\{ D\!\!\!\!\slash
_t[A]^{-1}\left[\left(\gamma_5\phi-i\eta\right) D\!\!\!\!\slash
_t[A]+  D\!\!\!\!\slash
_t[A]\left(\gamma_5\phi+i\eta\right)\right]\right\}.
\end{equation}

The trace operation is ill defined, and needs to be regularized.
We define our regularized $\omega$ as

\begin{equation}
\omega(t)_R=\lim_{M\rightarrow\infty}\tr\left\{ D\!\!\!\!\slash
_t[A]^{-1}\left[\left(\gamma_5\phi-i\eta\right) D\!\!\!\!\slash
_t[A]+  D\!\!\!\!\slash
_t[A]\left(\gamma_5\phi+i\eta\right)\right]e^{-R/M^2}\right\}.
\end{equation}
The choice of $R$ is always dictated by physical considerations.
For instance, if we are considering a gauge theory, we must take
into account regularization prescriptions which do not spoil gauge
invariance at the quantum level. This is usually achieved by
taking $R= D\!\!\!\!\slash _t[A]^2$, where $A_\mu$ is the gauge
field. Here, the model under study is not a gauge theory and
therefore we have even more freedom to choose the regulator. We
shall employ an operator of the form $R= D\!\!\!\!\slash _t[B]^2$,
where $B_\mu$ is certain vector field to be determined. We can
write $\omega(t)_R$ as
$\omega(t)_R=\omega_0(t)+\omega_\text{nc}(t)$ where

\begin{align}
\omega_0(t)=&\tr\left(2\gamma_5\phi e^{-R/M^2}\right)\\
\omega_\text{nc}(t)=&\tr\left\{ D\!\!\!\!\slash
_t[A]^{-1}\left(\gamma_5\phi-i\eta\right) \left[ D\!\!\!\!\slash
_t[A],e^{-R/M^2}\right]\right\}.
\end{align}
Here the subscript $0$ indicates the term that we would have
obtained if we had employed the cyclic property of the trace in
equation (\ref{omega2}). The subscript $\text{nc}$ refers to a
``non cyclic" term (this issue is discussed in detail in ref.
\onlinecite{CS}). The final expressions for these two terms are

\begin{align}
\omega_0(t)&=-(1-t)\frac{g}{\pi}\int
d^2x\,\phi\epsilon_{\mu\nu}\partial_\mu B_\nu\\
\omega_\text{nc}(t)&=-(1-t)\frac{g}{2\pi}\int
d^2x\,\partial_\mu(B_\nu-A_\nu)(\epsilon_{\nu\mu}\phi+\delta_{\nu\mu}\eta).
\end{align}

At this point it is a matter of straightforward algebra to check
that taking

\begin{align}
B_0=&A_0\\
B_1=&-A_1
\end{align}
we arrive at the same results obtained in the previous section
(equations (\ref{resultJacobian})-(\ref{Mmatrix})). Thus, we have
found an explicit form for the regulating operator of a Jacobian
associated to a non covariant fermionic determinant. This form, in
turn, leads to the correct dispersion relation for the
Tomonaga-Luttinger model. This is our main result. Let us mention
that the non covariant Jacobian given by eq.
(\ref{resultJacobian}) has been employed as an ansatz in previous
works on functional bosonization of Luttinger liquids
\cite{previous}. The derivation of this Jacobian was one of the
principal motivations of the present work.

\section{Bosonized Hamiltonian and Currents}

Up to this point we have worked in the Lagrangian formulation, but
in condensed-matter applications the Hamiltonian framework is
frequently preferred. It is then desirable to show how to derive,
in the functional bosonization framework discussed in this paper,
the usual Hamiltonian for the 1D electronic system, i.e. the
bosonic form of the Tomonaga-Luttinger model \cite{reviews}. The
other point we address in this section is the bosonic form of the
original fermionic currents (charge-density and electrical
current) and the issue of conservation.

Taking into account the expression for $\det D\!\!\!\!\slash\;[A]$
calculated in the preceding sections, and the relation between the
$\phi$ and $\eta$ fields and the $A_\mu$ fields
(eq.(\ref{AphietaRelation})), we can express the generating
functional (\ref{generatingFunctional}) in terms of the $A_\mu$
field:

\begin{multline}
\Z[S]=\N\int\D A_\mu\exp\left(-\frac{1}{2}\int
d^2x\,d^2y\,A_\mu(x)D_{\mu\nu}(x-y)A_\nu(y)\right)\times\\\exp\left(
-\frac{1}{2 g^2}\int d^2x\,d^2y\, S_\mu(x) V_{(\mu)}^{-1}(x-y)
S_\mu(y)\right)\exp\left(-\frac{1}{g}\int d^2x\,d^2y\,A_\mu(x)
V_{(\mu)}^{-1}(x-y) S_\mu(y)\right),
\end{multline}
where $D_{\mu\nu}$ is given in Fourier space by

\begin{equation}
D_{\mu\nu}(p)=\frac{g^2}{\pi(p_0^2+\vf^2 p_1^2)}
\begin{pmatrix}
    \vf p_1^2 & p_0p_1\\
    p_0 p_1  & -\vf p_1^2 \\
\end{pmatrix} +
\begin{pmatrix}
    \frac{1}{V_0} & 0 \\
    0 & \frac{1}{V_1} \\
\end{pmatrix}.
\end{equation}

We can decouple the $A_\mu$ field from the source $S_\mu$ by the
usual procedure of performing a translation in the $A_\mu$ field:

\begin{equation}
A_\mu\rightarrow A_\mu + \frac{D_{\mu\nu}^{-1}S_\nu}{gV_{(\nu)}},
\end{equation}
obtaining

\begin{multline}
\Z[S]=\N\int\D A_\mu\exp\left(-\frac{1}{2}\int
d^2x\,d^2y\,A_\mu(x)D_{\mu\nu}(x-y)A_\nu(y)\right)\times\\\exp\left[\frac{1}{2}\int
d^2x\,d^2y\,S_\mu(x)\Delta_{\mu\nu}^{-1}(x-y)S_\nu(y)\right]
\end{multline}
where $\Delta_{\mu\nu}^{-1}$ is given in Fourier space by

\begin{equation}
\Delta_{\mu\nu}^{-1}(p)=\frac{1}{\pi(p_0^2+v^2 p_1^2)}
\begin{pmatrix}
    -Kvp_1^2 & p_0p_1\\
    p_0 p_1  & \frac{v}{K}p_1^2 \\
\end{pmatrix},
\end{equation}
and the stiffness constant $K$ is given by

\begin{equation}
K=\sqrt{\frac{\vf-g^2V_1/\pi}{\vf+g^2V_0/\pi}}.
\end{equation}
We can multiply and divide by

\begin{equation}
\int\D A_\mu\exp\left(-\frac{1}{2}\int
d^2x\,d^2y\,A_\mu(x)\Delta_{\mu\nu}(x-y)A_\nu(y)\right),
\end{equation}
and perform an additional translation in the $A_\mu$ field

\begin{equation}
A_\mu\rightarrow A_\mu + \Delta_{\mu\nu}^{-1} S_\nu,
\end{equation}
to obtain

\begin{equation}
\Z[S]=\tilde{\N}\int\D A_\mu \exp\left(-\frac{1}{2}\int
d^2x\,d^2y\,A_\mu(x) \Delta_{\mu\nu}(x-y) A_\nu(y)+\int
d^2x\,S_\mu(x)A_\mu(x)\right).
\end{equation}
Finally, by defining the fields $\varphi$ and $\theta$ in the
following way:

\begin{align}
A_0&=\frac{-1}{\sqrt{\pi}}\,\partial_x\varphi\\
A_1&=\frac{i}{\sqrt{\pi}}\,\partial_x \theta,
\end{align}
we end up with the following generating functional

\begin{multline}
\Z[S]=\bar{\N}\int\D\varphi\D\theta\exp\left(-\frac{1}{2}\int dx
d\tau\left[\frac{v}{K}(\partial_x\varphi)^2+v K
(\partial_x\theta)^2+2i\partial_x\theta\partial_\tau\varphi\right]\right)\times\\
\exp\left(\int dx d\tau\,\left[-S_0\partial_x\varphi/\sqrt{\pi} +
i S_1\partial_x\theta/\sqrt{\pi}\right]\right).
\end{multline}

We then naturally identify the $\varphi$ field with the charge
density mode of the system and $\Pi=\partial_x\theta$ as its
canonical conjugate field. Moreover, the first two terms in the
quadratic action of the previous expression can be identified with
the Hamiltonian of the system:

\begin{equation}
\H=\frac{1}{2}\int dx\,\left[\frac{v}{K}(\partial_x\varphi)^2+v K
(\partial_x\theta)^2\right],
\end{equation}
which exactly coincides with the Hamiltonian obtained using
standard operational bosonization \cite{reviews}. Now, by
functional derivation we get the bosonic form of the currents

\begin{align}
j_0&=\frac{-1}{\sqrt{\pi}}\,\partial_x\varphi\\
j_1&=\frac{i}{\sqrt{\pi}}\,\Pi,
\end{align}
which, of course, are identical to the ones obtained in the
operator approach. It is important to stress that these currents
do not obey the continuity equation. Following ref.(17), one
introduces a physical electric current $j$, which is in general
different from $j_1$. The charge density is identified with $j_0$
($j_0=\rho)$. The physical current is determined by demanding that
the conservation law is verified:

\begin{equation}
\frac{\partial\rho}{\partial\tau} + \frac{\partial j}{\partial
x}=0.
\end{equation}
We obtain

\begin{equation}
j=\frac{i}{\sqrt{\pi}}\,v\,K\,\Pi.
\end{equation}
Note that only for $V_1=0$ ($g_2=g_4$ in the Tomonaga-Luttinger
language) one has $v\,K=1$ and $j=j_1$. As explained in ref.(17),
this difference between $j$ and $j_1$ is due to the fact that, in
general, the density does not commute with the interactions.

\section{Conclusions}
We considered a fermionic determinant associated to a non
covariant field theory. In particular we studied the determinant
which arises when implementing a path-integral approach to
bosonization based on the decoupling of the fermionic determinant
through appropriate changes of variables in the functional
integration measure. The model analyzed in this work (a non
covariant version of the Thirring model) has been previously used
to describe 1D highly correlated electronic systems which display
the so called Luttinger liquid behavior.

In the context of the heat-kernel regularization method, by
exploiting the freedom originated in the non covariance, we
determined a regulating operator that yields a bosonic action
which leads to the general form (in terms of the various coupling
constants) for the dispersion relations. These dispersion
relations are in full agreement with the ones that are well-known
in the operational framework. Previous path-integral computations
had used a covariant regularization, borrowed from relativistic
field theory, which gives a correct spectrum only for particular
values of the coupling constants.

We showed how to derive the bosonized Hamiltonian and currents,
which coincide with the ones obtained through standard operational
bosonization. In this way we were able to establish the precise
heat-kernel regularization that yields complete agreement between
the path-integral and operational approaches to the bosonization
of the Tomonaga-Luttinger model.

\setcounter{equation}{0}

\section*{Acknowledgements}

This work was partially supported by Universidad Nacional de La
Plata, Consejo Nacional de Investigaciones Cient\'{\i}ficas y
T\'ecnicas, CONICET and Fundaci\'on Antorchas.

\end{document}